\documentclass[sigconf]{acmart}
\AtBeginDocument{%
  \providecommand\BibTeX{{%
    \normalfont B\kern-0.5em{\scshape i\kern-0.25em b}\kern-0.8em\TeX}}}

\setcopyright{acmlicensed}
\copyrightyear{2018}
\acmYear{2018}
\acmDOI{XXXXXXX.XXXXXXX}

\acmConference[MSR 2024]{21st International Conference on Mining Software Repositories}{April 2024}{Lisbon, Portugal}

\acmISBN{978-1-4503-XXXX-X/18/06}

\usepackage{graphicx}
\usepackage{todonotes}
\begin{document}

\title{SciCat: A Curated Dataset of Scientific Software Repositories}


\author{Addi Malviya-Thakur}
\affiliation{%
  \institution{University of Tennessee}
  \institution{Oak Ridge National Laboratory}
  \city{Knoxville}
  \country{USA}
  }
\email{amalviya@vols.utk.edu}

\author{Reed Milewicz}
\affiliation{%
  \institution{Sandia National Laboratories}
  \city{Albuquerque}
  \country{USA}}
\email{rmilewi@sandia.gov}


\author{Lavínia Paganini}
\affiliation{%
  \institution{Eindhoven University of Technology}
  \city{Eindhoven}
  \country{The Netherlands}}
\email{l.f.paganini@tue.nl}

\author{Ahmed Samir Imam Mahmoud}
\affiliation{%
  \institution{University of Tartu}
  \city{Tartu}
  \country{Estonia}}
\email{ahmed.imam.mahmoud@ut.ee}

\author{Audris Mockus}
\affiliation{%
  \institution{University of Tennessee}
  \city{Knoxville}
  \country{USA}}
\email{audris@utk.edu}

 \begin{abstract}

The proliferation of open-source scientific software for science and
research presents opportunities and challenges. 
 In this paper, we introduce the SciCat dataset— a comprehensive collection of Free-Libre Open Source Software (FLOSS) projects, designed to address the need for a curated repository of scientific and research software.
This collection is crucial for understanding the creation of scientific software and aiding in its development.
To ensure extensive coverage, our approach involves selecting projects from a pool of 131 million deforked repositories from the World of Code data source. Subsequently, we analyze README.md files using OpenAI's advanced language models. Our classification focuses on software designed for scientific purposes, research-related projects, and research support software.
The SciCat dataset aims to become an invaluable tool for researching science-related software, shedding light on emerging trends, prevalent practices, and challenges in the field of scientific software development. Furthermore, it includes data that can be linked to the World of Code, GitHub, and other platforms, providing a solid foundation for conducting comparative studies between scientific and non-scientific software.
 \vspace{-.05in}

\end{abstract}

\ccsdesc[500]{Human-centered computing~Empirical studies in collaborative and social computing}
\ccsdesc[500]{Software and its engineering}

\keywords{Open-Source Software, Scientific Repositories, Software Classification, Large Language Models, Mining Software Repositories
}


\maketitle

\section{Introduction}
\vspace{-.05in}
Computing is an indispensable tool for science and engineering, and the demand for computational science and engineering (CSE) software has grown significantly. The CSE community has embraced modern software engineering practices and open-source, community-driven approaches to software development. This change has accelerated discovery, encouraged collaboration, and improved transparency and reproducibility in science \cite{yasar2003elements}. However, despite these advances, there remains a challenge in organizing these large collections of Free-Libre Open Source Software (FLOSS) projects, particularly in grouping them according to various characteristics relevant to scientific inquiry.

Our paper aims to address this gap by developing a large and
representative corpus of open-source scientific software~\cite{mama2021}. Using
World of Code, a massive collection of FLOSS repositories and Open
AI's advanced large-language models, we classified repositories
based on their README.md files thus creating a curated collection of
science- and research-related software repositories. The primary objective
of the resulting dataset is to enable studies that focus on trends, common practices, challenges, and evolution in scientific software development.

This corpus will enable a better understanding of how scientific software is maintained and sustained compared to non-scientific software. It also provides a comprehensive view of the software development landscape across various disciplines, institutions, and countries, with the intention of significantly contributing to the understanding and advancement of scientific software development. To the best of our knowledge, this is the first time such a dataset is presented to the empirical software engineering community interested in studying the scientific software ecosystem.

The following sections of this paper will detail our related work, methodology, description of the dataset and impact, limitations, and outline plans for future work, followed by our conclusion.
 \vspace{-.05in}
\begin{figure*}
\vspace{-.15in}
    \centering
    \includegraphics[width=1\linewidth]{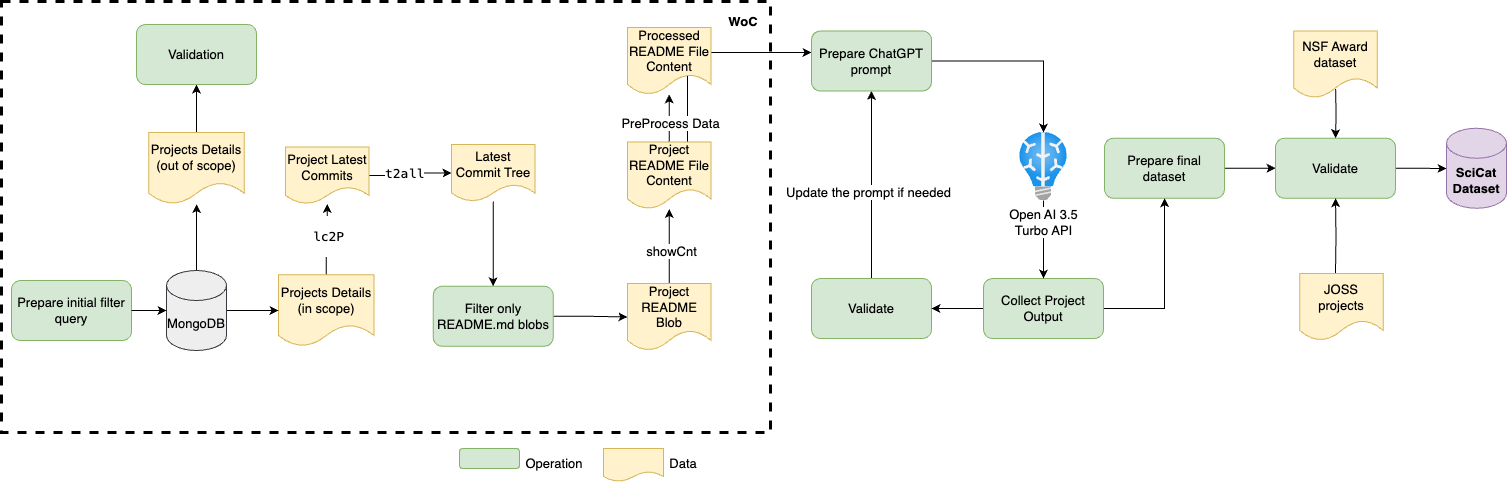}
\vspace{-.25in}
    \caption{Workflow showing the steps taken for obtaining SciCat Dataset}
    \label{fig:enter-label}
\vspace{-.15in}
\end{figure*} 
\vspace{-.18in}
\section{Related Work}
\vspace{-.05in}
As a class of software, scientific software has been relatively understudied by software engineering researchers, and existing studies that have compiled datasets have largely (1) been limited to small, targeted subsets of projects or (2) focused on data sources other than software repositories specifically. Murphy et al.~\cite{murphy2020curated} constructed a dataset focused on security defects in 20 popular scientific software projects. Milewicz et al. performed a study of contributor roles in 7 selected projects\cite{milewicz2019characterizing}. Mao et al. demonstrated a framework for metadata extraction from scientific software repositories using a corpus of 70 projects\cite{mao2019somef}. Building on that work, Kelley and Garijo\cite{kelley2021framework} developed an approach for constructing a knowledge graph of scientific software metadata based on 10,000 entries on Zenodo; we note that many of the repositories mentioned in that dataset are one-off repositories for storing data and scripts associated with particular papers. Meanwhile, Trujillo et al.\cite{trujillo2022penumbra} created a dataset of open-source software hosted outside of GitHub and Gitlab, including (but not limited to) academic projects hosted on university servers. Finally, Brown et al.~\cite{brown2023softsearch} introduced the Soft-Search datasets, which infer software production activities from grant documentation of NSF-funded projects from 2010 to 2023; this dataset ---while highly useful in other ways--- does not present links for GitHub repositories of the projects or a form to check the software developed by the award recipient. 
\vspace{.009in}

\vspace{-.07in}
\section{Methodology}
\subsection{Defining Scientific Software}
 \vspace{-.05in}
For the purposes of this study, we define scientific application software using the definition provided by Kelly: "application software that includes a large component of knowledge from the scientific application domain and is used to increase science knowledge for the purpose of solving real world problems"~\cite{KELLY201550}. In the taxonomy of research software established by Sochat et al.~\cite{sochat2022research}, this includes "software to directly conduct research" at the top of the software stack (e.g., physics simulations, brain imaging tools for neuroscience, etc.) while excluding more general-purpose software that directly or indirectly supports that research (e.g., math libraries, visualization tools, notebooks and workflows, scientific package managers)~\cite{mmcma2021}.
\vspace{-.20in}
\subsection{Data Source: World of Code}
 \vspace{-.05in}
In our study, we adopt a systematic, automated approach to identify and categorize scientific application software within the FLOSS ecosystem. Our primary data source is the World of Code (WoC) collection, a vast and dynamic repository of software projects. The prototype of the WoC infrastructure~\cite{10.1007/s10664-020-09905-9} was developed to facilitate the creation of theoretical, computational, and statistical frameworks. These frameworks are designed to uncover, collect, and analyze FLOSS operational data, thus constructing FLOSS supply chains. In addition, they help identify and quantify risks and develop effective practices and tools for risk mitigation. The WoC collection is rich with cross-referenced data, including information on authors, projects, commits, blobs, dependencies, and historical trajectories of software projects.

The WoC database is regularly updated and includes billions of git objects, which serves as an invaluable resource for our research. For the purposes of our study, we used version V of the WoC data, which was updated with new repositories found between March 1-30, 2023 and git objects retrieved by mid-May 2023; it includes approximately 3.93 billion commits, 15.43 billion file trees, 16.25 billion blobs, and 209.05 million projects (repositories) and 131.17 million deforked projects. More details on the WoC dataset, 
can be found on the World of Code website\footnote{\url{https://worldofcode.org/}}.

 \vspace{-.05in}
\subsection{Constructing a Cross-Section of FLOSS Repositories using World of Code}
\vspace{-.05in}
We aim to identify and focus on scientific software that is not only developed for a single document publication but is also actively maintained and updated.  
We first made a query meant to identify salient repositories, based on the approach of Carruthers et al. \cite{carruthers2022software}. We considered the number of files in the repository (at least 5), number of commits (more than 300), number of contributors to the repository (at least 2), the number of months with activity on the project (at least 6 consecutive active months), the latest commit date (more recent than November 2018), and the programming language(s) identified in the repository (to remove repositories without any identifiable code). Through this process, we filtered through 131 million repositories in the WoC database and downselected 430,469 projects that met our criteria.

Among these, 423,596 projects were identified as containing a README file in their root directory, a primary focus of our analysis. We then applied a series of filters to ensure data consistency and relevance. The first filtering step involved the exclusion of repositories with inconsistent naming conventions, which refined our dataset to 422,221 projects. Our analysis then narrowed to projects with README files specifically in the Markdown format (".md"), resulting in a subset of 354,943 projects. In the context of our research, the README.md file serves as a key informational interface, analogous to a web landing page; we hypothesized that scientific software packages would provide us with adequate detail in their READMEs to allow us to classify them correctly. Subsequent stages of our data processing involved the identification and removal of duplicated records, characterized by multiple occurrences of the same project name and blob. This step reduced the dataset to 349,992 projects. After eliminating projects with duplicate names, we were left with a final count of 349,652 unique projects.

Following these filtration steps, we extracted the content of the README files from the WOC, culminating in a collection of 349,650 README files. An additional inclusion of 658 projects was made post-deduplication process. The final dataset, comprising 350,308 README.md files, was meticulously compiled and prepared for the application of OpenAI's scientific classification methodologies. Among these, 342,656 projects were successfully classified by the OpenAI process and then joined the metadata and curated the final datasets. 
Subsequent to data collection, we embarked on a two-fold processing strategy. First, we extracted the project metadata from WoC MongoDB. These metadata include information about the project's contributors, development timeline, and other pertinent details. Second, we processed the collected readme texts using OpenAI's Turbo 3.5 API
for analyzing the content of the readmes.
 \vspace{-.09in}
\subsection{Using a Large Language Model (LLM) to ask designed questions about README content}
 \vspace{-.04in}
We focused on scientific application software. Our reasoning is that asking the LLM to judge whether a repository is a piece of application software based on the README is easier to check, whereas the further out you go from the application layer (I/O libraries, visualization tools, package managers, operating systems, etc.) there is a lot of software that hypothetically helps answer a scientific question.

Our research specifically targets scientific application software within the FLOSS ecosystem.
We defined the terms explicitly and then sent a refined prompt to the Open AI's gpt 3.5 turbo model.
We define scientific application software as those developed for specific scientific or engineering purposes, excluding more general software tools and libraries. To accurately categorize and evaluate software projects, we utilized prompt engineering techniques, following the best practices outlined in 'A Prompt Pattern Catalog to Enhance Prompt Engineering with ChatGPT~\cite{openai_chatgpt}.

The methodology involved testing our approach on approximately 100 repositories. We use ChatGPT to analyze the readmes and categorize each project based on its content and our predefined criteria. The analysis focused on several key aspects:
\vspace{-.05in}
\begin{itemize}
    \item Determining if a project is a scientific application software.
    \item Identify any mentions of research publications, papers, or funding in the readme.
    \item Classifying the software as research software.
    \item Recognizing whether the project falls under the category of science support software.
    \item Describing the purpose of the software project in one or two sentences.
    \item List 5-6 keywords or phrases that best describe the software's domain.
\end{itemize}

\vspace{-.07in}
\subsection{Validation of dataset}
\vspace{-.05in}
\begin{table}[]
\vspace{-.05in}
\caption{Description of Data Fields in the dataset}
\vspace{-.15in}
\label{tab:cc}
\resizebox{\columnwidth}{!}{%
\begin{tabular}{|l|l|}
\hline
\textbf{Field Name} & \textbf{Description} \\ \hline
ProjectID & A unique identifier for each project. \\ \hline
url & A column for the URL of the project, likely on GitHub. \\ \hline
isScientificAppSoftware & A Boolean column indicating whether the project is scientific application software. \\ \hline
mentionsPaperOrFunding & Indicates if the project mentions a paper or funding. \\ \hline
isResearchSoftware & A Boolean field to specify if the project is research software. \\ \hline
isScienceSupportSoftware & Indicates whether the software supports scientific research. \\ \hline
keywords & A list of keywords associated with the project. \\ \hline
shortdescription & A brief description of the project. \\ \hline
fullanswer & A longer, more detailed description or answer related to the project. \\ \hline
NumForks & Number of forks of the project repository. \\ \hline
NumStars & Number of stars in the project repository. \\ \hline
NumFiles & Number of files in the project repository. \\ \hline
NumAuthors & Number of Authors for the project repository. \\ \hline
AuthorsGender & Provides information on the gender distribution of the project's authors. \\ \hline
NumCommits & The total number of commits made in the project. \\ \hline
NumBlobs & The number of blobs (binary large objects) in the project. \\ \hline
CodeLanguages & Lists the programming languages used in the project. \\ \hline
NumActiveMon & The number of active months of the project. \\ \hline
LatestCommitDate & The date of the last commit. \\ \hline
EarliestCommitDate & The date of the earliest commit. \\ \hline
CommunitySize & Size of the community involved in the project. \\ \hline
blobId & Unique identifier for the blob in the project. \\ \hline
commitID & Unique identifier for commit in the project. \\ \hline
\end{tabular}%
}
\vspace{-.15in}
\end{table}
For validation,
we examined 60 randomly selected projects using two distinct approaches: Cohen's Kappa and F1 score metrics. The evaluations were conducted by two independent raters, and for the F1 metrics, one rater's evaluations served as a 'pseudo ground truth' due to the absence of a ground truth dataset. For Q1, "Is it scientific application software?" Cohen's Kappa between OpenAI's API and the primary rater was 0.150, indicating slight agreement. The Kappa value between the two human raters was slightly lower at 0.130. The F1 score metrics
showed
a Precision of 0.4, Recall of 0.857, Accuracy of 0.653, and an F1 Score of 0.545. For Q2, "Does the Readme mention paper or funding?" The Kappa value between OpenAI's API and the primary rater was significantly higher at 0.667, suggesting substantial agreement. In contrast, the agreement between the two human raters was almost perfect, with a Kappa of 0.892. The F1 score metrics
revealed
a Precision of 0.733, Recall of 0.917, Accuracy of 0.755, and a higher F1 Score of 0.815. For question 4, "Is it science support software?" the Cohen's Kappa between OpenAI's API and the primary rater was -0.146, implying less than chance agreement, while the human-human rater comparison yielded a Kappa of 0.342, indicating fair agreement. The F1 score metrics
indicated
a Precision of 0.185, Recall of 0.625, Accuracy of 0.347, and a significantly lower F1 Score of 0.286.
The collective findings
from our analysis,
highlight differing degrees of concurrence between OpenAI's API and human assessors. This assessment reveals variations in agreement levels among the questions,
 providing insights
performance across different question types. In particular, when we consider the primary purpose of our dataset, the metrics related to Q2 strongly support the identification of projects related to science and research.
\begin{scriptsize}
\begin{table}[b]
\vspace{-.05in}
\caption{Number of projects in dataset based on different criteria}
\vspace{-.15in}
\label{table:projects_criteria}
\begin{tabular}{|l|c|p{7.9cm}|}
\hline
\textbf{Criteria} & \textbf{\# of Projects} \\ \hline
Scientific Application Software & 14455 \\ \hline
Mentions Paper or Funding & 16661 \\ \hline
Research Software & 5750 \\ \hline
Science Support Software & 186219 \\ \hline
Scientific App & 6349 \\
and Mentions Paper or Funding & \\ \hline
Research Software & 3323 \\
and Mentions Paper or Funding & \\ \hline
Research Software or Scientific App & 8085 \\
and Mentions Paper or Funding & \\ \hline
Total projects in final dataset & 342656 \\ \hline
\end{tabular}
\vspace{-.15in}
\end{table}
\end{scriptsize}

To further validate the dataset used in these assessments, we employed several strategies to ensure the dataset's robustness. First, we cross-referenced the dataset with the Journal of Open Source Software (JOSS) dataset, a recognized platform for peer-reviewed open source scientific software \cite{katz2018joss}. This comparison resulted in 445 projects overlapping, of which 173 were classified as Scientific Software, validating a
subset of the dataset. In addition, we used a validation approach based on the National Science Foundation (NSF) dataset. Although none of the NSF datasets contained actual repository names or URLs, we conducted manual sampling based on NSF grants credited in the README.md files of the repositories, ensuring consistency and identifying additional scientific software projects \cite{brown2023softsearch}. We also conducted a manual review. However, it is important to acknowledge the challenges of validating such a large dataset. Despite inherent limitations, these strategies provide a solid foundation for assessing the quality of the dataset.

\vspace{-.02in}
\section{ Dataset Description and Impact}
 \subsection{Description}
 \vspace{-.05in}
The provided dataset  includes several columns with various types of information. Each row in the dataset represents a unique project. The details of the columns can be seen in Table-\ref{tab:cc}.
Finally, our dataset is available online in the form of a Pickle file in a tar.gz archive file. It also includes codes to generate the dataset. \footnote{\url{https://anonymous.4open.science/r/SciCatMSRDataShowCase/}} and
summary statistics describing the dataset including number of projects found in different categories are shown in Table \ref{table:projects_criteria}. 
\vspace{-.07in}
 \subsection{Potential Research Questions }
\vspace{-.05in}


The SciCat dataset is a large-scale dataset of scientific software repositories that spans numerous disciplines, enabling researchers to study trends and patterns specific to libraries and ecosystems that contribute to science and research. 
Using this dataset, researchers can gain insight into a wide array of research questions on the development, sustainability, and impact of scientific software, such as collaboration patterns, software practices, effects of funding and publications on scientific software projects.
Some examples 
include:
\begin{itemize}
\item \textbf{RQ1}: To what extent does gender diversity permeate the software teams of the FLOSS scientific applications? How does this diversity manifest within the landscape of project size and team composition?

\item \textbf{RQ2}: What is the frequency of security vulnerabilities in FLOSS scientific application software projects? How can we craft a comprehensive portrayal of the security posture and methodologies adopted by these projects?

\item \textbf{RQ3}: What are common collaboration patterns in open-source scientific software projects, and how do they impact project outcomes?

\item \textbf{RQ4}: How can cross-disciplinary analyses benefit scientific software development, and what is the prevalence of essential libraries across different fields?

\item \textbf{RQ5}: To what extent are machine learning and AI techniques integrated into scientific software, and what is their impact?

\item \textbf{RQ6}: What are the key challenges and opportunities specific to scientific software development, and how can they improve usability, code quality, and overall software development practices?


\end{itemize}
\vspace{-.05in}
 These research questions are just a few examples of the myriad possibilities offered by the SciCat dataset. 
 
 

\section{Limitations and Future improvements} 
\vspace{-.05in}
Our study is subject to several significant limitations that require careful consideration. First, the accuracy and completeness of our data are heavily dependent on the information extracted from the README.md files
. Any inaccuracies or deficiencies in README.md files could result in misclassifications and omissions of critical project details. Second, our dataset is inherently influenced by selection bias, as it is primarily composed of projects available in the World of Code database and those actively maintaining README.md files. This bias may lead to an underrepresentation of projects on platforms other than popular ones and those that do not utilize README.md files. 
Another limitation pertains to our reliance on large language models (LLMs), specifically OpenAI's GPT-3.5 model. The accuracy of our results depends on the performance and limitations of this model. Furthermore, the lack of comprehensive datasets for validation and the impracticality of manually verifying such a large dataset adds to the challenges we face in ensuring the accuracy and representativeness of our findings.
We aimed to enhance the project sample size by beginning with a vast dataset of 131 million repositories, thereby improving dataset representativeness and reducing selection bias. Utilizing the World of Code (WoC) data source helps mitigate selection bias concerns associated with specific platforms, as it consolidates data from various open-source platforms, including Github, GitLab, SourceForge, Bitbucket, and more.
To further mitigate this limitation, in the future, we will make efforts to diversify the dataset, even more, using other smaller curated platforms such as
institutional repositories, improving the dataset to ensure that it better reflects the wider landscape of scientific software projects~\cite{arma2021}.  
To address the limitations posed by the use of LLMs and their performance, we plan to use the keywords collected from our initial analysis to further refine and validate the dataset with a subsequent analysis or modeling. 
 
 It is also important to acknowledge that certain limitations, particularly those related to the quality of the README.md file and the accuracy of the LLM, may persist and require ongoing efforts to improve. Hence, for future improvements, we would consider supplementing the README.md analysis with additional contextual information, such as project documentation, source code, and user feedback. In addition, our goal will be to raise awareness about the establishment of standardized guidelines for README.md, encouraging developers to provide comprehensive and accurate information, and encouraging the citation of articles and grants. 
 We will also explore alternative validation strategies, such as surveying project maintainers or users, to provide additional validation of the dataset, especially for projects with limited external references.

 \vspace{-.05in}
\section{Conclusion}
\vspace{-.04in}
The SciCat dataset not only offers a substantial opportunity to reshape our understanding of computational science and scientific software engineering, but also stands as a singular resource in its field. Currently, there are no other comprehensive datasets readily accessible to researchers focusing on mining software repositories and conducting empirical studies within the scientific software ecosystem. This absence underscores the unique and critical contribution of the SciCat dataset. By filling this gap, it becomes an indispensable tool for researchers aiming to dive into the nuances and dynamics of scientific software. Its distinctive position enhances its potential to provide unparalleled insights, making it a cornerstone for future explorations and developments in
scientific software research.

\balance

\bibliographystyle{ACM-Reference-Format}
\bibliography{references}

\end{document}